\documentclass[aps,prd,onecolumn,groupedaddress,showpacs,showkeys, nofootinbib,amssymb]{revtex4}
\usepackage[T1]{fontenc}
\usepackage[latin1]{inputenc}
\usepackage{graphicx}
\usepackage[english]{babel}
\usepackage{bm}
\usepackage{amsmath}
\usepackage{amssymb}
\usepackage{amsfonts}
\usepackage{epsfig}
\usepackage{colordvi}
\usepackage{color}

\begin{document}

\title{Noether symmetry of FRW cosmology with  f - essence}
\author{Tolkynay Myrzakul$^{1}$\footnote{Email address: trmyrzakul@gmail.com}, Ertan G\"udekli$^{2}$\footnote{Email address: gudekli@istanbul.edu.tr},
Shynaray Myrzakul $^{1}$\footnote{Email address: srmyrzakul@gmail.com}, 
Ratbay Myrzakulov$^{3}$\footnote{Email address: rmyrzakulov@gmail.com}}

\affiliation{$^1$Kazakh National University, Almaty, 050040, Kazakhstan.}
\affiliation{$^2$Departament of Physics, Istanbul University, Istanbul, Turkey.}
\affiliation{$^3$Eurasian International Center for Theoretical Physics and Department of General Theoretical Physics, Eurasian National University, Astana 010008, Kazakhstan,}

\begin{abstract}
In this paper, we investigate the dynamics of the universe on the background of f-essence when a non-minimal coupling with gravity. Field equations are obtained and using the Noether theorem, explicit forms of the coupling function and the Lagrange function of the f-essence were obtained. Accordingly, for this model, the cosmological parameters that describe the accelerated expansion of the universe are determined.
 \end{abstract}
\maketitle

\tableofcontents

%%%%%%%%%%%%%%%
\section{Introduction}

In recent years, due to the rapid development of astrophysics and the possibilities of modern technology, a huge amount of astronomical data has been obtained, such as supernova type Ia \cite{Riess, Perlmutter}, cosmic microwave background (CMB) anisotropy \cite{WMAP}, baryon acoustic oscillations \cite{Eisentein}, weak lensing \cite{weakl} and Large Scale Structure \cite{LSS}. Due to this, a fundamental discovery was made, which showed that our Universe is expanding with acceleration. On the basis of this discovery, many models appeared that describe the dynamics of the universe as a whole, and as well as its various localized objects. The main theoretical interpretation that describes the accelerated expansion of the universe is the so-called "dark energy" \cite{DE1, DE2, DE3, DE4, DE5, DE6}. As is known the general theory of relativity describes the  dynamics of the universe from the point of view of classical field theory. However, this theory can not explain the physics of the early and modern Universe. To fill this gap, many modifications of this theory have been constructed, such as the quintessence \cite{quin}, the brane cosmology \cite{brane}, the k-essence \cite{k-essence, k-essence1, k-essence2, k-essence3, k-essence4, k-essence5, k-essence6}, the f-essence \cite{f-essence1, f-essence2}, the g-essence \cite{g-essence1, g-essence2, g-essence3} and others.

In this paper, we will consider one of these modified theories for the homogeneous and isotropic Friedman-Robertson-Walker space-time (FRW), a cosmological model of the f-essence, which is a particular case of the r-essence and the fermion analogue of the c-essence (for more details, see \cite{f-essence1}). The authors of these works, studying the f-essence as a gravitational source of accelerated expansion, have shown that the fermion field, initially having an anisotropic space, becomes isotropic, forming singularities of free cosmological solutions that describe well the accelerated expansion of the Universe. Difference of our article from the above works is that we generalized the Lagrangian to the case of a minimal coupling of f-essence with a gravitational field. Using this Lagrangian, we obtain a system of field equations, and also use the Noether theorem to determine the explicit forms of the coupling function and the Lagrange function of the f-essence.

The famous theorem of the German mathematician Emma Noether, published in 1918 \cite{Noether}, makes it possible to analyze the studied physical system on the basis of the available symmetry data that this system possesses. The theorem compares the number of continuous symmetries of the considered system to the number of conservation laws, that is, the number of conserved quantities, which are called conserved or Noether charges. The mathematical advantage of using this method in physical problems, and in particular in astrophysics and cosmology, consists in the fact that this method makes it possible to simplify the system of differential equations determining the dynamics of the considered physical system, and also to determine the integrability of the physical system itself. The Noether method of symmetry also allows us to verify the self-consistency of the studied physical model. 

Earlier, the approach of Noether's theorem in cosmology was considered in \cite{Noether1, Noether2, Noether3, Noether4, Noether5}. For example, Noether gauge symmetry for F (R) gravitation was studied in \cite{Kucu} under the Palatini formalism, which determines the form of  $f(R)$ and obtains an exact solution for the cosmological scale factor. The authors of \cite{Rudi} used this approach for the fermion field. This model also determined the evolution of the scale factor, which is the main factor that describes the dynamics of the universe. In \cite{Noether1} solutions are obtained that can describe the phantom accelerated expansion of the universe. 

The paper consists of six sections. In the first section of the main part, we present f-essence   non-minimally coupled to the $R$ gravity  and field equations derived from it for spatially flat metric FRW. Next, in the second section, we search the Noether symmetry approach for the point-like Lagrangian derived in the first section. The cosmological solutions are analyzed for Ansatz f-essence derived in third section. At last section we present the final remarks and conclusions. 

We use units of $k_{\mathrm{B}} = c = \hbar = 1$ and 
$8\pi/M_{Pl}^2=1$, where $M_{Pl}$ is the Planck Mass. 

%%%%%%%%%%%%%%%

\section{$R$ gravity model with  f-essence}
In this section we consider a spatially flat FRW Universe within the f-essence field that is non-minimally coupled to gravity in the framework of $R$ gravity. Action for this model written as
\begin{eqnarray}
\label{act1} 
S  = \int d^4  x e \left[h(u) R+2K(Y,u)\right],
\end{eqnarray}
where $R$  are curvature scalar,  $\psi$ and $\bar{\psi} = \psi^{\dag}\gamma^{0}$ denote the spinor field and its adjoint, respectively, the dagger represents complex conjugation, $u=\bar{\psi}\psi$ is the bilinear function, $Y$ - the canonical kinetic term of the fermionic field, $h(u)$ is a generic function, representing the coupling of gravity with f-essence field, $K$ is the Lagrangian of the f-essence field depending of $Y$ and $u$.

We will consider here the simplest homogeneous and isotropic cosmological model, FRW, whose spatially flat metric is given by

\begin{eqnarray} \label{FRW}
ds^2  = -dt^2 +  a^2(t) \left(dx^2 + dy^2 + dz^2\right),
\end{eqnarray}
where $a(t)$ is the scale factor of the Universe. For this metric, the vierbein is chosen to be $(e^\mu_a)=diag(1,1/a,1/a,1/a)$ and $(e^a_\mu)=diag(1,a,a,a)$. Also we have
\begin{eqnarray}
R=6 \left( \frac{{\ddot a}}{a}+\frac{{\dot a}^2}{a^2}\right), \  Y=\frac{1}{2}i\left(\bar{\psi}\gamma^0\dot{\psi}-\dot{\bar{\psi}}\gamma^0\psi\right).
\end{eqnarray}
where
\begin{equation}
\gamma^0 = \begin{pmatrix} I & 0 \\ 0 & -I \end{pmatrix}, \quad\gamma^k = \begin{pmatrix} 0 & \sigma^k \\ -\sigma^k & 0 \end{pmatrix},\quad \gamma^5 = \begin{pmatrix} 0 & I \\ I & 0 \end{pmatrix},
\end{equation}
are the Dirac gamma matrices and an overdot denotes derivative with respect to proper time $t$.
 
Then the above action becomes
\begin{eqnarray}\label{act2}
S&=&\int d^4x \left(6 a^2 \ddot{a} h + 6 a \dot{a}^2 h + 2a^3K\right).
\end{eqnarray}
where $a$  spatially flat FRW  spacetime has been adopted. After an integration by parts, the point-like Lagrangian assumes the following form

\begin{eqnarray}\label{PointLagra1}
 L&=&6 a \dot{a}^2 h + 6 a^{2} \dot{a} h_{u} \dot{u}  - 2a^3K,
\label{Lag}\end{eqnarray}
here, because of homogeneity and isotropy of the metric it is
assumed that the spinor field depends only on time, i.e. $\psi=\psi(t)$.

Then, taking into account $H = \dot{a}/a$ denotes the Hubble parameter, we have equations of motion

\begin{equation}
3 H^{2} h + 3 H \dot{h} - \rho_f = 0,
\label{EM1}
\end{equation}
\begin{equation}
\left(2 \dot{H} +3 H^2\right) h +   \ddot{h} + 2 H\dot{h}  + p_f = 0,
\label{EM2}
\end{equation}

\begin{eqnarray}
K_{Y} \dot{\psi}+0.5\left(3 H+\dot{K_Y}\right)\psi-i\left[K_u+3\left(\dot{H}+2H^2\right)h_u\right]\gamma^0\psi=0,
\label{EM3}
\end{eqnarray}
\begin{eqnarray}
K_{Y} \dot{\bar{\psi}}+0.5\left(3 H+\dot{K_Y}\right)\bar{\psi}+i\left[K_u+3\left(\dot{H}+2H^2\right)h_u\right]\bar{\psi}\gamma^0=0,
\label{EM4}
\end{eqnarray}

Also from the 	consequent of Lorentz-invariance of the energy-momentum tensor we have conservation law in next form
\begin{equation}
\dot{\rho_f} + 3 H \left(\rho_f + p_f\right) = 0,
\end{equation}
here  $p_f=K$ and $\rho_f = Y K_{Y} - K$ are the energy density and pressure of the f-essence field \cite{f-essence1}. From the equations (\ref{EM3})-(\ref{EM4}) we obtain
\begin{eqnarray}
\ u&=& \frac{u_0}{a^3K_Y},
\label{u}
\end{eqnarray}
where $u_0$ is a constant. As we see the field equations (\ref{EM1}) - (\ref{EM4}) are non-linear equations of a high order. As we know, the solution  of these equations are very hard. In the next section, for solve these equations we will use the well-known approach used in physics Noether's symmetry.

%%%%%%%%%%%%%%%%%%%%%%%%%%%%%%%%
\section{The Noether Symmetry Approach}
\label{quattro}
In this section we will be detail consider approach the Noether symmetry. This symmetry play in modern physics important role. It is applied to determine the conserved quantities and constants of motion.  The  Noether symmetry approach tells us that Lie derivative of the Lagrangian with respect to a given vector field ${\bf X}$ vanishes, i.e.
\begin{equation}
{\bf X} L = 0. \label{noether}
\end{equation}  

We will search the Noether symmetries for our model. In terms of the components of the spinor
field $\psi= \left(\psi_{0}, \psi_{1}, \psi_{2}, \psi_{3}\right)^{T}$ and its adjoint $\bar{\psi} =
\left(\psi^{\dagger}_{0}, \psi^{\dagger}_{1}, - \psi^{\dagger}_{2}, -\psi^{\dagger}_{3}\right)$, the point-like Lagrangian (\ref{Lag}) rewrite as
\begin{eqnarray}\label{PointLagra11}
 L&=& 6a \dot{a}^2h + 6 a^{2} \dot{a} h_{u} \sum^3_{i=0}\left(\dot{\psi}^{\dagger}_{i} \psi_{i}+\psi^{\dagger}_{i} \dot{\psi}_{i}\right) - 2 a^{3} K,\label{lag1}
\end{eqnarray}
above $\bf X$ is defined for present dynamics
\begin{eqnarray} 
{\bf X}=\alpha \frac{\partial}{\partial a}+{\dot \alpha} \frac{\partial}{\partial  \dot a} + \sum^{3}_{j=0}\left(\beta_{j} \frac{\partial}{\partial \psi_{j}}+{\dot \beta_{j}}  \frac{\partial}{\partial\dot \psi_{j}} + \gamma_{j}  \frac{\partial}{\partial \psi^{\dagger}_{j}}+{\dot \gamma_{j}}  \frac{\partial}{\partial \dot \psi^{\dagger}_{j}}\right).
\label{ourX}
\end{eqnarray}
Here
\begin{eqnarray} 
{\dot \alpha}\,&=&\, \frac{\partial \alpha}{\partial a}{\dot a}+ \sum^3_{j=0}\left(\frac{\partial\alpha }{\partial \psi_{j}}\dot{\psi_{j}}+\frac{\partial\alpha}{\partial \psi^{\dagger}_{j}}\dot{\psi^{\dagger}_{j}}\right),\\
{\dot \beta_{j}}\,&=&\,\frac{\partial \beta_{j}}{\partial a}{\dot a}+\sum^3_{j=0}\left( \frac{\partial \beta_{j}}{\partial \psi_{j}}\dot{\psi_{j}}+\frac{\partial \beta_{j}}{\partial \psi^{\dagger}_{j}}\dot{\psi^{\dagger}_{j}}\right),\\
{\dot \gamma_{j}}\,&=&\, \frac{\partial \gamma_{j}}{\partial a}{\dot a}+\sum^3_{j=0}\left( \frac{\partial \gamma_{j}}{\partial \psi_{j}}\dot{\psi_{j}}+\frac{\partial \gamma_{j}}{\partial \psi^{\dagger}_{j}}\dot{\psi^{\dagger}_{j}}\right),
\end{eqnarray}
where $\alpha,\beta_i$ and $\gamma_i$ are unknown functions of the variables $a, \psi_{i}$ and $\psi_{i}^\dagger$. 

Then applying condition (\ref{noether}) for equation (\ref{lag1})  we obtain an equation that explicitly depend on $\dot{a}^{2},\dot{\psi}^{2}_{i},\dot{\psi}^{\dagger 2}_{i},\dot{a} \dot{\psi}_{i}, \dot{a} \dot{\psi}^{\dagger}_{i}, \dot{a}, \dot{\psi}_{i}, \dot{\psi}^{\dagger}_{i}$ and equating to zero the coefficients of the data variables have the following system of differential equations

\begin{equation} \label{not1}
\alpha + 2 a \frac{\partial \alpha}{\partial a} + a \frac{h_{u}}{h} \sum_{i=0}^{3} \left(\epsilon_{i} \beta_{j} \psi^{\dagger}_{i} + \epsilon_{i} \gamma_{j} \psi_{i}\right) + a^{2} \frac{h_{u}}{h} \sum_{i=0}^{3} \left(\frac{\partial \beta_{j}}{\partial a} \psi^{\dagger}_{i} + \frac{\partial \gamma_{j}}{\partial a} \psi_{i}\right) = 0,
\end{equation}
\begin{equation}\label{p1}
6 a^{2} h_{u} \psi^{\dagger}_{i} \frac{\partial \alpha}{\partial \psi_{j}} = 0,
\end{equation}
\begin{equation}\label{p2}
6a^{2} h_{u} \psi_{i} \frac{\partial \alpha}{\partial \psi^{\dagger}_{j}} = 0,
\end{equation}
\begin{equation} \label{not2}
\left(2 \alpha + a \frac{\partial \alpha}{\partial a}\right) h_{u} \psi^{\dagger}_{i} + 2 h \frac{\partial \alpha}{\partial \psi_{j}} + a h_{u} \sum_{i=0}^{3} \left(\frac{\partial \beta_{j}}{\partial \psi_{j}} \psi^{\dagger}_{i} + \frac{\partial \gamma_{j}}{\partial \psi_{j}} \psi_{i}\right) + a \left(\beta_{j} (h_{u})_{\psi_{j}} + \gamma_{j} (h_{u})_{\psi^{\dagger}_{j}}\right) \psi^{\dagger}_{i} + \gamma_{j} a h_{u} = 0,
\end{equation}
\begin{equation} \label{not3}
\left(2 \alpha + a \frac{\partial \alpha}{\partial a}\right) h_{u} \psi_{i} + 2 h \frac{\partial \alpha}{\partial \psi^{\dagger}_{j}} + a h_{u} \sum_{i=0}^{3} \left(\frac{\partial \beta_{j}}{\partial \psi^{\dagger}_{j}} \psi^{\dagger}_{i} + \frac{\partial \gamma_{j}}{\partial \psi^{\dagger}_{j}} \psi_{i}\right) + a \left(\beta_{j} (h_{u})_{\psi_{j}} + \gamma_{j} (h_{u})_{\psi^{\dagger}_{j}}\right) \psi_{i} + \beta_{j} a h_{u} = 0,
\end{equation}
\begin{equation} \label{not4}
\sum_{i=0}^{3} \left(\frac{\partial \alpha}{\partial \psi_{j}} \psi_{i} + \frac{\partial \alpha}{\partial \psi^{\dagger}_{j}} \psi^{\dagger}_{i}\right) = 0,
\end{equation}
\begin{equation} \label{not5}
\sum_{i=0}^{3} \left(\frac{\partial \beta_{j}}{\partial a} \psi^{\dagger}_{i} - \frac{\partial \gamma_{j}}{\partial a} \psi_{i}\right) = 0,
\end{equation}
\begin{equation}\label{not6}
3 \alpha \psi^{\dagger}_{j} + a \gamma_{j} + a \sum_{i=0}^{3} \left(\frac{\partial \beta_{j}}{\partial \psi_{j}} \psi^{\dagger}_{i} - \frac{\partial \gamma_{j}}{\partial \psi_{j}} \psi_{i}\right) = 0,
\end{equation}
\begin{equation}\label{not7}
3 \alpha \psi_{j} + a \beta_{j} - a \sum_{i=0}^{3} \left(\frac{\partial \beta_{j}}{\partial \psi^{\dagger}_{j}} \psi^{\dagger}_{i} - \frac{\partial \gamma_{j}}{\partial \psi^{\dagger}_{j}} \psi_{i}\right) = 0,
\end{equation}
\begin{equation} 
3 \alpha \left(K - Y K_{Y} \right) +  a K_{u} \sum_{i=0}^{3} \left(\epsilon_{i} \beta_{i} \psi^{\dagger}_{i} + \epsilon_{i} \gamma_{i} \psi_{i}\right) = 0.\label{n3}
\end{equation}
In this system we introduce the symbol
\begin{equation} 
 \epsilon_i=\begin{cases}+1 \qquad\hbox{for}\qquad i=1,2,\cr
 -1 \qquad\hbox{for}\qquad i=3,4.\end{cases}
\end{equation}

From the equations (\ref{p1}), (\ref{p2})  we see that function $\alpha$  obviously do not dependent on the variables $\psi_i, \psi^{\dagger}_i$. It's only  function of $a$. The equation (\ref{n3}) rewrite as
\begin{equation} 
\frac{3 \alpha \left( K - Y K_{Y}\right)}{a K_{u}} = - \sum_{i=0}^{3} \left(\epsilon_{i} \beta_{i} \psi^{\dagger}_{i} + \epsilon_{i} \gamma_{i} \psi_{i}\right).\label{not8}
\end{equation}

If put this equation in expressions (\ref{not1})-(\ref{not4})  and %учитывая  
that functions $h$ and $K$ depend on the variables $u$ and $Y$ we get the following result
\begin{equation}
\label{alpha}
\alpha = \alpha_{0} a^{n},
\end{equation}
where $\alpha_{0}$ and $n$ are some constants. Also, from the equations (\ref{not4})-(\ref{not6}) and (\ref{not8}) we also can defines symmetry generators $\beta_{j}$ and $\gamma_{j}$ as
\begin{equation}
\beta_{j} = - \left(\frac{3}{2} \alpha_{0} a^{n-1} + \epsilon_{j} \beta_{0}\right) \psi_{j},\label{beta}
\end{equation}
\begin{equation}
\gamma_{j} = - \left(\frac{3}{2} \alpha_{0} a^{n-1} - \epsilon_{j} \beta_{0}\right) \psi^{\dagger}_{j},\label{gamma}
\end{equation}
where $\beta_{0}$ is a constant. Using the solutions (\ref{beta}),(\ref{gamma}) and (\ref{alfa}) and equation (\ref{not8}) we find the functions $K$ as
\begin{equation}
K=\mu Y - \nu u,\label{f-lag}
\end{equation}
where $\mu$, $\nu$ are an integrable constant. Here $T=\mu Y$ is a kinetic term and $V=\nu u$ is a potential term for the Lagrangian fermion field. Finally, after some algebraic calculations, we find the solution for the coupling function $h(u)$ as 
\begin{equation}
h(u) = h_{0} u^{\frac{1+2n}{3n}},\label{nonm}
\end{equation}
here $h_{0}$ is an integration constant. If $n=0$ or $n=-\frac{1}{2}$ the coupling function in the equation (\ref{nonm}) is equal to 1, then in the action (\ref{act2})  the interaction between of gravity and fermionic field will be minimal. In section IV, we will use the values for the coupling function $h(u)$ is equation (\ref{nonm}) and Lagrangian fermionic field (\ref{f-lag}) for the determination of the cosmological solutions our model. 

At the end of this section, we would like to determine the charge Noether for our model
\begin{equation}
Q = \alpha \frac{\partial L}{\partial \dot{a}} + \sum_{i=0}^{3} \left(\beta_{i} \frac{\partial L}{\partial \dot{\psi}_{i}} + \gamma_{i} \frac{\partial L}{\partial \dot{\psi}^{\dagger}_{i}}\right) = const.
\end{equation}

Then we have
\begin{equation}
Q = \tilde{Q}a^b\dot{a} -2i\eta_0 u_0, 
\end{equation}
where
\begin{equation}
\tilde{Q} =  -12 \alpha_1h_1\frac{1+q}{q}\left(\frac{u_0}{K_1}\right)^\frac{1+2q}{q},\ \ b=\frac{q^2-q-1}{q}  
\end{equation}
are constants and we have next solution for scale factor
\begin{equation}
a=a_0(t-t_0)^\frac{1}{b+1}
\end{equation}

%%%%%%%%%%%%%%%%%%%%%%%%%%%%%%%%
\section{Cosmological solutions}
\label{quattro}
In this section, for describe the dynamics of the universe, we try analytical solve the field equations (\ref{EM1}) - (\ref{EM4}). For this, we need to find the explicit dependence of the scale factor $a$ of the time $t$. 

First case, we consider the case when $n=-\frac{1}{2}$, then the coupling function $h=h_0$ and %предположим что 
$h_0=1$ (see equation (\ref{nonm})). Then, the equations (\ref{EM1}) - (\ref{EM4})  we rewrite as following 
\begin{equation}
3 H^{2} -\rho_f = 0, \label{EM_1}
\end{equation}
\begin{equation}
3 H^{2} + 2 \dot{H} + p_f = 0,
\label{EM_2}
\end{equation}
\begin{eqnarray}
K_{Y} \dot{\psi}+0.5\left(3 H+\dot{K_Y}\right)\psi-i K_{u} \gamma^{0} \psi=0,
\label{EM_3}
\end{eqnarray}
\begin{eqnarray}
K_{Y} \dot{\bar{\psi}}+0.5\left(3 H+\dot{K_Y}\right)\bar{\psi}+i K_{u}\bar{\psi}\gamma^{0}=0.
\label{EM_4}
\end{eqnarray}
\begin{equation}
\dot{\rho_f} + 3 H \left(\rho_f + p_f\right) = 0,
\label{EM_5}
\end{equation}
where $\rho_f = Y K_{Y} - K$ and $p_f=K$. Earlier,  in the papers [12]-[16] considered various 
cosmological solutions of the field equations (\ref{EM_1})-(\ref{EM_5}). Also, from the equations (\ref{u}) and (\ref{f-lag}) we also find
\begin{equation}
u = \frac{u_{0}}{\mu a^{3}}.
\end{equation}

If put the equations (34) and (41) in equation (36),  we obtain 
\begin{equation}
a = \left(\frac{3 \nu u_{0}}{4 \mu}\right)^{\frac{1}{3}} \left(t - t_{0}\right)^{\frac{2}{3}},
\end{equation}
where $t_{0}$ is a constant of integration. The Hubble parameter is
\begin{equation}
H= \frac{2}{3 \left(t - t_{0}\right)}.
\end{equation}

Then, for our model the energy density and pressure are
\begin{equation}
\rho = \frac{4}{3 \left(t-t_{0}\right)^{2}}, \ p=0.
\end{equation}

And the equation of state parameter for this model have the next form
\begin{equation}
\omega= 0.
\end{equation}
One of the main cosmological parameter is the deceleration parameter. If $q<0$ indicates the standart decelerating models and if$q>0$ corresponds to accelerating models. When $q=0$ is expansion with a constant velocity.
\begin{equation}
q = -\frac{\ddot{a}}{a H^2}= -\frac{1}{2n},
\end{equation}
These solutions describe a standard pressureless matter field.

Second case, we consider the general case when the coupling function is equal to $h=h_{0} u^{\frac{1+2n}{3n}}$. For this case, the scalar factor is obtained 
\begin{equation}
a(t) = a_{0} \left(t-t_{0}\right)^{\frac{2n}{n-1}},
\end{equation}
where $a_{0} = \left(\frac{4 n^{3} \nu u_{0}^{\frac{n-1}{3n}}}{3 \left(1-n\right) \left(n+1\right)^{2}h_{0}} \right)^{\frac{n}{n-1}}$. 

The corresponding Hubble parameter is 
\begin{equation}
H= \frac{2n }{\left(n-1\right) \left(t-t_{0}\right)},
\end{equation}

From fermion field are energy density and pressure follows from field equations (7) and (8) as
\begin{eqnarray}
\rho_f = \rho_0\left(t-t_0\right)^\frac{6 n}{1-n}\,
\rho_0=\frac{12 n h_0}{1-n}\left(\frac{u_0}{\mu a_0^3}\right)^\frac{1+2 n}{3 n}
\end{eqnarray}
and
\begin{equation}
p_f =  \frac{\rho_0 }{6 n \left(n-1\right)}\left(t-t_0\right)^{\frac{6 n}{1-n}}\left(8 n^2 +3 n +1 - \left(t-t_0\right)^{-1}\right),
\end{equation}

Then, we can define the equation of state parameter for the fermion field by equations (52)-(53) as 
\begin{equation}
\omega = \frac{p}{\rho} = \omega_0 -\frac{\omega_1}{t-t_0}, 
\end{equation}

where $\omega_0=\frac{8 n^2 +3 n+ 1}{6 n (n+1)}$ and $\omega_1=\frac{8 n^2 }{6 n (n+1)}$.

Also, finally this section we find the deceleration parameter is found similarly to the first case 
\begin{equation}
q = -\frac{\ddot{a}}{a H^2}= -\frac{1}{2}-\frac{1}{2 n}=const,
\end{equation}

%%%%%%%%%%%%%%%%%%%%%%%%%%%%%%%%%%%%%%%%%%%%%%%%%%%%%%%%%%%%%%%%%%%%%%%%%%
\section{Final remarks and conclusion}
%%%%%%%%%%%%%%%%%%%%%%%%%%%%%%%%%%%%%%%%%%%%%%%%%%%%%%%%%%%%%%%%%%%%%%%%%%

This paper investigates the evolution of the accelerated expansion of the late-time Universe on the background of the f-essence. The work begins by considering action of the field in FRW flat space-time,  taking into account its non-minimally coupling with f-essence (1). The equations of motion in the form of equations (7) - (10) have obtained. To find field-matter coupling and the explicit form of the f-essence Lagrangian used Noether symmetry (14) condition and three sets of solutions (equation (31), (32) and (33)).

Using the approach Noether symmetry is because this approach can reduce the number of unknown variables of a dynamic system, which leads to the possibility of solving the equations of motion of the theory. In addition, this approach can be considered as a physically motivated criterion, since such symmetry are always related to conserved quantities. Such conserved quantity of our system or so called Noether charge defined in the form (37) using equation (36).

Finally, in the fifth chapter cosmological parameters of  f-essence studied in detail. Two cases, when $n = -1 / 2$ and for any other values of $n$ have been considered. In the first case it is shown that the coupling of the field with the matter does not change with time, the scale factor evolves, as shown in equation (46), similarly to dust-like matter of the standard cosmological model. Parameter condition in this case is equal to zero (49), a deceleration parameter (50) is positive, which is the standard setting state baryonic matter. For the second case, the scale factor evolves in the form of (51) as n increases acceleration scale factor tends to a linear function of time. For $n = 0$, the scale factor remains unchanged and field-matter coupling becomes infinite, however when $n = 1$ non-minimal coupling of the field with  matter becomes linear, and the scale factor tends to infinity.

\end{document}